\documentclass[useAMS,usenatbib]{mn2e}

\usepackage{url,times,graphicx,amsmath,amsfonts,amssymb,aas_macros,color,epsfig,epstopdf}

\newcommand{\hMpc}{{\ifmmode{h^{-1}{\rm Mpc}}\else{$h^{-1}$Mpc}\fi}}
\newcommand{\hkpc}{{\ifmmode{h^{-1}{\rm kpc}}\else{$h^{-1}$kpc}\fi}}
\newcommand{\hMsun}{{\ifmmode{h^{-1}{\rm {M_{\odot}}}}\else{$h^{-1}{\rm{M_{\odot}}}$}\fi}}
\newcommand{\ltsima}{$\; \buildrel < \over \sim \;$}
\newcommand{\gtsima}{$\; \buildrel > \over \sim \;$}
\newcommand{\lsim}{\lower.5ex\hbox{\ltsima}}
\newcommand{\gsim}{\lower.5ex\hbox{\gtsima}}
\def\LCDM{$\Lambda$CDM}

\def\lesssim{\mathrel{\hbox{\rlap{\hbox{\lower4pt\hbox{$\sim$}}}\hbox{$<$}}}}
\def\gtrsim{\mathrel{\hbox{\rlap{\hbox{\lower4pt\hbox{$\sim$}}}\hbox{$>$}}}}

\newcommand{\Fig}[1]{Fig.~\ref{#1}}
\newcommand{\beq}{\begin{equation}}
\newcommand{\eeq}{\end{equation}}
\def\beqa{\begin{eqnarray}}
\def\eeqa{\end{eqnarray}}
\def\hMpc{$h^{-1}\,{\rm Mpc}$}
\def\hkpc{$h^{-1}\,{\rm kpc}$}
\def\LCDM{\ensuremath{\Lambda}CDM}
\def\LWDM{\ensuremath{\Lambda}WDM}
\def\Vmax{$V_{\rm max}$}
\def\Rmax{$R_{\rm max}$}

\def\head{
 \vbox to 0pt{\vss
                   \hbox to 0pt{\hskip 440pt\rm LA-UR-10-07069\hss}
                  \vskip 25pt}}

\title[Massive dark matter subhaloes of the Milky Way]
{Too small to succeed? Lighting up massive dark matter subhaloes of the Milky Way}
\author[Di Cintio et. al]
       {Arianna Di Cintio$^{1}$\thanks{E-mail: arianna.dicintio@uam.es}, Alexander Knebe$^{1}$, Noam I. Libeskind$^2$, Gustavo Yepes$^1$, \newauthor Stefan Gottl\"ober$^2$, Yehuda Hoffman$^3$\\
$^{1}$Departamento de F\'isica Te\'orica, M\'odulo C-15, Facultad de Ciencias, Universidad Aut\'onoma de Madrid, 28049 Cantoblanco, Madrid, Spain\\
$^2${Leibniz-Institut f\"{u}r Astrophysik} Potsdam, An der Sternwarte 16, D-14482 Potsdam, Germany\\
$^3$Racah Institute of Physics, The Hebrew University of Jerusalem, Givat Ram, Israel
}

\begin{document}

\date{Accepted XXXX . Received XXXX; in original form XXXX}

\pagerange{\pageref{firstpage}--\pageref{lastpage}} \pubyear{2010}

\maketitle

\label{firstpage}


\begin{abstract}
Using Constrained Local UniversE Simulations (CLUES) of the formation of the Local Group in a cosmological context we investigate the recently  highlighted problem that the majority of the most massive dark subhaloes of the Milky Way are too dense to host any of its bright satellites. In particular, we examine the influence of baryonic processes and find that they leave a twofold effect on the relation between the peak of the rotation curve and its position (\Vmax\ and \Rmax). Satellites with a large baryon fraction experience adiabatic contraction thus decreasing \Rmax\ while leaving \Vmax\ more or less unchanged. Subhaloes with smaller baryon fractions undergo a decrease in \Vmax\ possibly due to outflows of material. Furthermore, the situation of finding subhaloes in simulations that lie outside the confidence interval for possible hosts of the bright MW dwarf spheroidals, appears to be far more prominent in cosmologies with a high $\sigma_8$ normalisation and depends on the mass of the host. We conclude that the problem cannot be simply solved by including baryonic processes and hence demands further investigations.

\end{abstract}
\noindent
\begin{keywords}
 methods: numerical - $N$-body simulations -- galaxies: formation - haloes - Local Group
 \end{keywords}

\section{Introduction} \label{sec:introduction}
The $\Lambda$ Cold Dark Matter (\LCDM) model, first explored more than two decades ago \citep{Davis85}, has been very successful in explaining a multitude of observations at cosmological scales, such as anisotropies of Cosmic Microwave Background radiation (CMB) \citet[e.g.][]{Jarosik11} and galaxy clustering on large scales \citep[e.g.][]{Cole05}. However, on smaller, galactic scales the tests of the \LCDM\ model are complicated by the baryonic physics involved in galaxy formation. Therefore, testing the currently accepted concordance model at these scales is necessary in order to not only understand the nature of dark matter but also the accuracy of the model itself.

The validity of the \LCDM\ model on galactic scales is still being questioned due to the discrepancy between the number of observed satellites and  the number of predicted dark matter subhaloes. High resolution simulations of galactic-size haloes resolve a substantial number of substructures within the virial radius, as first pointed out by \citet{Klypin99} and \citet{Moore99}, and recently reviewed by \citet{Kravtsov10} and \citet{Bullock10}.

The most popular interpretation of this so-called "Missing Satellite Problem" requires that the smallest dark matter haloes are inefficient at forming stars \citep[e.g.][]{Bullock10,Kravtsov10}. Mechanisms such as early reionization of the intergalactic medium and supernovae feedback have been invoked to identify the halo mass scale where the galaxy formation starts to be inefficient \citep{Bullock00,Somerville02,Benson02}, partially solving the problem. Furthermore, the detection of satellites is most certainly biased because of current detection limits \citep{Tollerud08,Walsh09}.

There is yet another aspect of the satellite population that needs to be addressed: the mismatch between the predicted and inferred distribution of \Vmax\ values at the high-\Vmax\ end as recently highlighted by \citet{Boylan11}, where \Vmax\ measures the peak of the rotation curve of subhaloes. Using the Aquarius simulations \citep{Springel08} and the Via Lactea II simulation \citep{Diemand08} they found that the majority of the most massive subhaloes (i.e. the high-\Vmax\ objects) of the Milky Way are too dense to host any of its bright satellites. 

There are a number of ways in which this discrepancy may be resolved: the subhalo mass function of the Milky Way could be a statistical anomaly with respect to the \LCDM\ expectations \citep{Liu10,Guo11}, or the fundamental assumption that the luminosities of the satellites are not monotonically related to the mass of the subhaloes does not hold true. 

In response to the claims by \citet{Boylan11}, \citet{Lovell11} explored the possibility that warm rather than cold dark matter can provide a better match to the inferred distribution of satellite circular velocities. With a power spectrum suppressed at masses below $\sim10^{10}M_\odot$ (corresponding to a warmon mass of 2~keV), they found that a warm dark matter model naturally produces haloes that are less concentrated than their cold dark matter counterparts. 
The attempt to explain the evolution of small scale structures in the local universe with a \LWDM\ model was already presented in \citet{TikhonovGott09}.
However, this is only one possible solution to the problem.

 Baryonic processes will most certainly also affect the dark matter distribution. \citet{Blumenthal86} showed that dissipative baryons will lead directly to the adiabatic contraction of the halo increasing its central density, thus being a critical ingredient to determine subhalo properties. However, the possibility that the influence of baryons will lead to a flattening of the dark matter central density cusp (through dynamical friction of infalling substructures composed of dark matter \textit{and} baryons) has, for instance, been suggested by \citet{El-Zant01} and further studied in \citet{Romano-Diaz08}. Another way in which the haloes' density can be reduced is through sudden mass outflows that can alter substantially the central structure, as suggested by \citet{Navarro96b}. In a recent work of \citet{Parry11} this last scenario has been tested by following the evolution of one simulated satellite, with promising results. The same authors though also showed that the inclusion of baryons in simulations does not seem to have any correlation with the increase or decrease of the dark matter central density.

In this \textit{Letter} we directly address the issue of the \Vmax\ problem in \LCDM\ simulations by comparing two identical simulations with each other: one that is solely based upon dark matter physics and another incorporating all the relevant baryonic physics. These simulations form part of the CLUES project\footnote{\texttt{http://www.clues-project.org}}, in which the initial conditions are set by imposing constraints derived from observational data of the Local Group. The main feature of using constrained simulations is that it provides a numerical environment that closely matches our actual neighborhood.

\section{The simulations} \label{sec:simulation}

\subsection{Constrained Simulations of the Local Group}
\label{sec:clues}
We use the same simulations already presented in \citet{Libeskind10}, \citet{Libeskind10infall}, \citet{Knebe10a}, and \citet{Knebe11a} and refer the reader to these papers for a more exhaustive discussion and presentation of these constrained simulations of the Local Group that form part of the CLUES project. However, we briefly summarize their main properties here for clarity.

We choose to run our simulations using standard \LCDM\ initial
conditions, that assume a WMAP3 cosmology \citep{Spergel07}, i.e.
$\Omega_m = 0.24$, $\Omega_{b} = 0.042$, $\Omega_{\Lambda} = 0.76$. We
use a normalization of $\sigma_8 = 0.75$ and a slope of the
power spectrum of $n=0.95$. We used the treePM-SPH MPI code \texttt{GADGET2}
\citep{Springel05} to simulate the evolution of a cosmological box
with side length of $L_{\rm box}=64 h^{-1} \rm Mpc$. Within this box
we identified (in a lower-resolution run utilizing $1024^3$ particles)
the position of a model local group that closely resembles the real
Local Group \citep[cf.][]{Libeskind10}. This Local Group has then been
re-sampled with 64 times higher mass resolution in a region of $2
h^{-1} \rm Mpc$ about its centre giving a nominal resolution
equivalent to $4096^3$ particles giving a mass resolution of $m_{\rm
  DM}=2.1\times 10^{5}$\hMsun\ for the dark matter and $m_{\rm
  gas}=4.42\times 10^4$\hMsun\ for the gas particles. For more details
we refer to the reader to \citet{Gottloeber10}.

For this particular study we further use a gas dynamical SPH
simulation started with the same initial conditions, in which we additionally follow the feedback and star formation rules of
\cite{Springel03}: the interstellar medium (ISM) is modeled as a two
phase medium composed of hot ambient gas and cold gas clouds in
pressure equilibrium. The thermodynamic properties of the gas are
computed in the presence of a uniform but evolving ultra-violet cosmic
background generated from QSOs and AGNs and switched on at $z=6$
\citep{Haardt96}.  Cooling rates are calculated from a mixture of a
primordial plasma composition. No metal dependent cooling is assumed.
Cold gas cloud formation by thermal instability, star formation, the
evaporation of gas clouds, and the heating of ambient gas by supernova
driven winds are assumed to all occur simultaneously.
Note that the results presented through the paper will only refer to the specific SF/feedback model of \citet{Springel03}: other formalisms might lead to different conclusions, and will be addressed in a companion paper (Di Cintio et al., in preparation).

In addition we also have at our disposal a dark matter only CLUES simulation based upon the WMAP5 cosmology \citep{Komatsu09} whose details will be presented in a companion paper; here it suffices to know that this simulation has the same formal resolution as the WMAP3 one, and it has also been re-simulated within a sphere of $2h^{-1} \rm Mpc$ radius, i.e. the primary difference between the two simulations is merely the cosmology.

\subsection{The (Sub-)Halo Finding}
\label{sec:halofinding}
We used the MPI+OpenMP hybrid halo finder \texttt{AHF} (AMIGA halo finder) to identify haloes and subhaloes in our simulation\footnote{\texttt{AHF} is freely available from \url{http://popia.ft.uam.es/AMIGA}}. \texttt{AHF} is the successor of the \texttt{MHF} halo finder by \citet{Gill04a}, and a detailed description of its mode of operation is given in the code paper \citep{Knollmann09}. Note that \texttt{AHF} automatically (and essentially parameter-free) finds haloes, sub-haloes, sub-subhaloes, etc. As the two WMAP3 simulations started with the same initial conditions (apart from the baryons) we can match individual subhaloes in the DM only simulation with a ``sister'' subhalo in the SPH run \citep[see][]{Libeskind10}. In effect, this cross identification pairs subclumps at $z=0$ that originated from the same overdensity in the initial conditions.

\section{Results}\label{sec:results}
In order to most closely match the results presented by \citet{Boylan11} and not to be contaminated by numerical effects we limited the subhaloes used throughout the study to those within 300kpc from their respective host and more massive than $M_{\rm sub}>2\cdot10^8M_\odot h^{-1}$. We further stack the data for the two most massive hosts representing our MW and M31 galaxies.

\begin{figure}
  \includegraphics[width=3in]{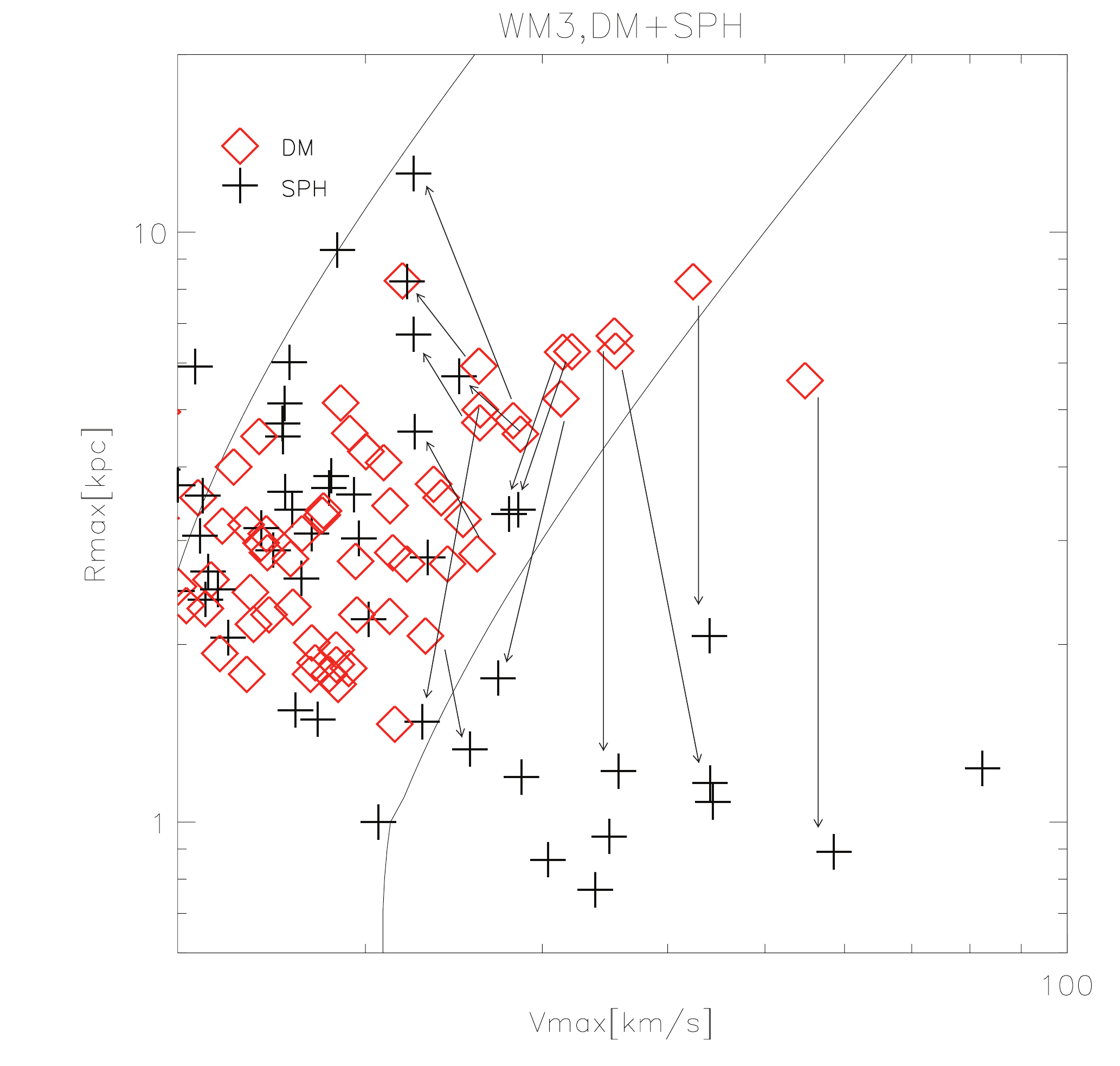}
  \caption{The relation between the peak of the rotation curve \Vmax\ and its position \Rmax\ for the WMAP3 simulations: the diamonds are DM only subhaloes, the cross represent baryonic SPH subhaloes. The two solid lines delimite the $1\sigma$ confidence interval of the observed bright Milky Way dwarf spheroidal galaxies, as in \citet{Boylan11}. The arrows connect the DM-SPH sister pairs found following the matching haloes procedure of \citet{Libeskind10}.}
\label{fig:WM3_DMGAS}
\end{figure}

\begin{figure}
  \includegraphics[width=3in]{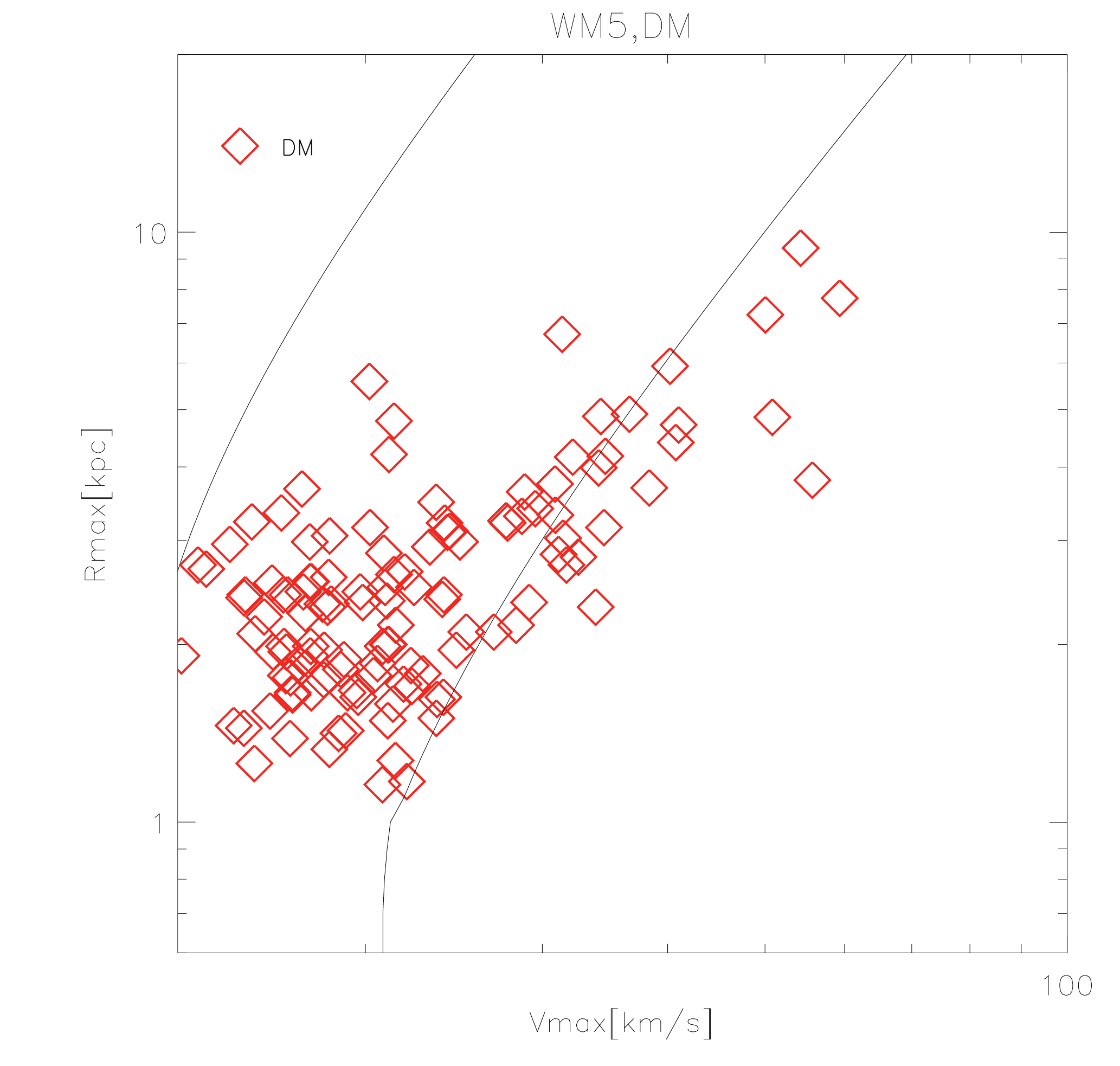}
  \caption{The same as \Fig{fig:WM3_DMGAS} but for the WMAP5 (dark matter only) simulation.}
\label{fig:WM5_DM}
\end{figure}

In \Fig{fig:WM3_DMGAS} we show the relation between \Rmax\ and \Vmax\ for the WMAP3 simulation alongside the $1\sigma$ confidence region of the known Milky Way satellites, assuming that the mass density profile of the subhaloes containing the nine observed dwarf spheroidal follows a NFW profile \citep{Navarro96}: the two solid lines in \Fig{fig:WM3_DMGAS} (and \ref{fig:WM5_DM}) thus limit the area consistent with the observed half-light radii and masses of these dwarfs, based on the work of \citet{Wolf10}. The diamonds represent the subhaloes in the dark matter only simulation while the crosses are the satellite galaxies in the SPH run. The lines connect the sister haloes, i.e. those objects that could be cross-identified in the two simulations. Please note that not all subhaloes could be cross-identified and hence only a certain number of them are connected by arrows. The results for the WMAP5 (dark matter only) data are presented separately in \Fig{fig:WM5_DM}.

The results of these plots are quite interesting. First of all we notice that massive subhaloes (i.e. high-\Vmax\ objects) appear to be outside the observational range only in the case of the WMAP5 cosmology. This is thus in agreement with the findings of \citet{Boylan11} whose Fig.2 shows both the Aquarius and Via Lactea II simulation data combined. The latter run used a $\sigma_8=0.74$ which is close to our WMAP3 value. Note that for this simulation the subhaloes are found only marginally outside the observational confidence interval. However, note that the actual \Vmax\ values for the subhaloes depend on the host mass and $V_{\rm max, host}$, respectively \citep[cf.][]{Reed05,Diemand07,Springel08}. Therefore, in order to better compare the WMAP5 to the WMAP3 simulation, we scaled the subhaloes' maximum velocities, $V_{\rm max,sub}^{WM5}$, by the ratio $V_{\rm max,host}^{WM3}/V_{\rm max,host}^{WM5}$ (not presented here) where the respective values are $V_{\rm max,MW}=131$, $V_{\rm max,M31}=128$ for WMAP3, and $V_{\rm max,MW}=178$, $V_{\rm max,M31}=194$ for WMAP5 (all in km/s). We find that this re-normalization leads to a $\approx 30$\% decrease of the  $V_{\rm max,sub}^{WM5}$ values, bringing them into agreement with the WMAP3 results. In that respect, the two dark matter only simulations are in fact not too different! 

More importantly, we see in \Fig{fig:WM3_DMGAS} that the inclusion of baryonic physics does not solve the problem of the massive and highly concentrated dark matter subhaloes. On the contrary subhaloes with baryons appear to be down-shifted in the \Rmax-\Vmax\ plane with respect to their dark matter counterpart, sometimes even entering the regime outside the observational constraints only in the SPH run. However, we also find that the lower-\Vmax\ objects seem to be shifted in the direction anticipated by \citet{Boylan11}, i.e. to the upper left of the plot. There appear to be two competing effects moving subhaloes in the \Rmax-\Vmax\  plane.

The six SPH (sister) subhaloes that are outside the confidential range have a smaller \Rmax\ than their DM only companion: the addition of baryons causes a contraction of the halo. This effect is also visible for three SPH (sister) subhalo inside the observational area and is readily explained by the physical phenomenon of adiabatic contraction \citep{Blumenthal86, Gnedin04}. We confirm that the baryon fraction $f_b=\Omega_b/\Omega_m$ of those subhaloes moving downwards is higher than for the subhaloes shifted to the upper left. On average, the baryon fraction of the nine (sister) SPH subhaloes, whose \Rmax\ is reduced with respect to their DM counterpart, is $f_b/f_{b,\rm cosmic}\sim0.314$, while the mean $f_b$ of the SPH subhaloes inside the $1\sigma$ area whose \Rmax\ increases is $f_b/f_{b,\rm cosmic}\sim0.006$, i.e. substantially smaller.  The subhaloes with high $f_b$ experience adiabatic contraction and the majority of these objects are the ones with the initial highest  \Rmax\ -- \Vmax\ pairs.

To confirm this last point, we used the \texttt{CONTRA} code \citep{Gnedin04} to calculate the adiabtic contraction of a dark matter halo in response to condensation of baryons. Using our numerical data as input parameters, we found that adiabatic contraction is actually efficient only for those subhaloes with sufficiently high $f_b$, as expected:  the amount of the \Rmax\ reduction computed this way perfectly matches the observed shifts in \Fig{fig:WM3_DMGAS}.

Instead, for the lower \Vmax\ sister subhaloes (with substantially smaller baryon fractions) the baryonic matter has the capability to lower the maximum velocity of the rotation curves, while increasing \Rmax. This has already been claimed in previous works and may be due to different mechanisms. In particular, we like to highlight the mass outflow model of \citet{Navarro96b}: immediate expulsion of a large fraction of baryonic material during star formation could be the cause of the creation of a central dark matter core, which will move the peak of the rotation curve to larger radii. This model has been successfully tested by \citet{Parry11} who followed the formation history of a single stellar dominated satellite, which undergoes the sequence of events predicted by \citet{Navarro96b}. Another possible explanation to end up with less concentrated density profiles, is through the mechanism described by \citet{Mashchenko06}. A random bulk motion of gas, driven by stellar feedback, results in a flattening of the central DM cusps, thus leading to DM densities smaller than predicted by pure DM cosmological models. But why is it that those objects with low baryon fractions are the ones that require the aforementioned mechanisms? Is it that the gas expulsion has already occurred, thereby lowering the baryon fraction? Possibly the baryon fraction is only low at redshift $z=0$ because of mass losses during the subhaloes history? Lately, \citet{Nickerson11} explored the effect of several baryon loss mechanisms on subhaloes in SPH simulation of a Milky Way like galaxy, too: they found that for the subhaloes which ended up having (or having had) stars but no gas the most efficient mechanism of baryons removal is exactly the stellar feedback \citep{Dekel86}. Finally, we note that the adiabatic contraction (following \citet{Gnedin04}) is ineffective for these subhaloes. We will address all these issue of the temporal evolution, mass loss and baryon influence in greater detail in a companion paper (Di Cintio et al., in preparation).

\begin{figure*}
  \includegraphics[width=7in]{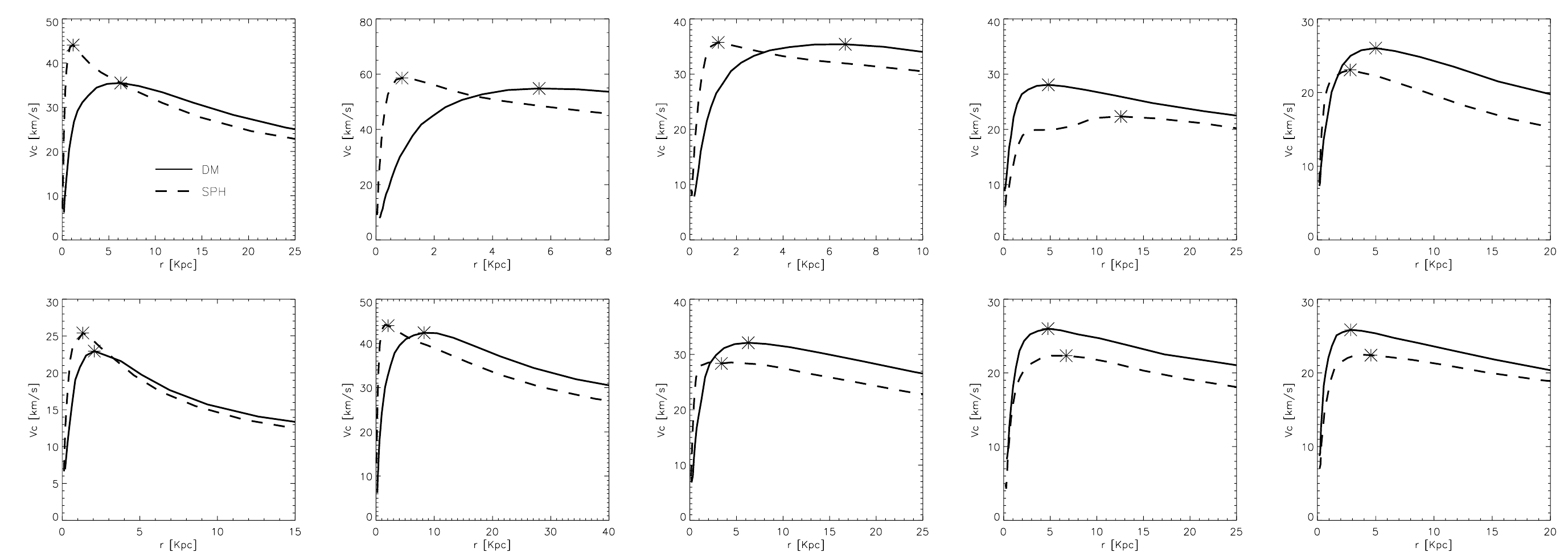}
  \caption{Rotation curves of ten sister pairs of massive subhaloes. In each panel the velocity profile of a pair of DM and SPH subhaloes is presented. The actual values of \Vmax-\Rmax\ are plotted as asterisks.}
\label{fig:rotcurve}
\end{figure*}

We close this Section with a detailed look at the rotation curves of the sister haloes in \Fig{fig:rotcurve}. In each plot the two sister objects are presented; the solid and dashed lines represent the circular velocity of the DM subhaloes in the dark matter only simulation and of the (sister) SPH subhalo, respectively. The asterisks show the \Vmax-\Rmax\ pairs used in \Fig{fig:WM3_DMGAS}. We thus observe adiabatic contraction at work: the first three objects (which happen to have high baryon fraction) in that plot clearly show the centrally peaked total matter distribution in the SPH run. The plot further indicates that our measurements of the rotation curve and its peak are not contaminated by numerical artifacts (e.g. mis-identified halo centre, etc.).

\section{Conclusions}\label{sec:conclusion}

In this letter we explored the possibility that baryonic processes may solve the recently presented problem of "the puzzling darkness of Milky Way subhaloes" \citep{Boylan11}. To this extent, we used dark matter only as well as full hydrodynamical simulations of cosmic structure in the context of the CLUES project. We used cosmological parameters determined from both the WMAP3 and WMAP5 data.

Our conclusions are twofold and can be summarized as follows:

\begin{itemize}

\item
We find that when baryonic physics is included, following the feedback and star formation prescriptions of \cite{Springel03}, the problem of having too dense massive subhaloes is not solved. Instead, gasdynamical simulations pose new questions regarding which mechanisms are responsible for the lowering of \Rmax\ in those subhaloes (while \Vmax\ remains more or less constant). Adiabatic contraction seems to be a reasonable explanation, as shown using the modified adiabatic contraction model of \citet{Gnedin04}: this process is effective only for some subhaloes, specifically, for those with a high baryon fraction. For the SPH subhaloes with lower baryon fractions at redshift $z=0$, instead, we observe a general increase of \Rmax\ with respect to their DM counterpart, thus meaning that other effects are at work, e.g. the model proposed by \citet{Navarro96b} in which a rapid expulsion of baryonic mass during star formation causes a reduction of the halo concentration, as well as naturally explaining the low baryon fraction of these objects.

\item
While in the WMAP5 DM only case we find dark matter subhaloes outside the confidence area (calculated following the prescription given in \citet{Boylan11}) in the WMAP3 cosmology we only have one massive subhalo outside this observational range. Since the Via Lactea II and Aquarius simulations presented in \citet{Boylan11} are similar cases, we conclude that the cosmology certainly has an influence, too: the higher $\sigma_8$ of the WMAP5 scenario eventually led to higher host masses which -- according to our test -- are the most likely reason for the higher number of excessively centrally concentrated substructures. Note that the latest data from WMAP7 favours $\sigma_8=0.807$, a value between the WMAP3 and WMAP5 results: this could mean that the problem is worse than in WMAP3, but not as pronounced as in the WMAP5 case.
\end{itemize}

An issue neither touched upon by us nor other authors is the adequacy of using NFW profiles when calculating the confidence interval for possible hosts of the bright MW dwarf spheroidals. It is obvious that tidal effects will lead to severe modifications of the original NFW density profile subhaloes had upon infall into their host \citep{Kazantzidis04}. They therefore leave an impact upon internal and kinematical properties, respectively \citep{Lokas10,Lokas11}, which should be taken into account when using observed half-light radii $R_{1/2}$ and their corresponding masses $M_{1/2}$ to define the confidence interval. Further, \citet{Romano-Diaz08} showed that adiabatic contraction makes the dark matter profile almost isothermal. However, the relevance is questionable as material will primarily be stripped from the outer regions: \citet{Penarrubia08} state that dSphs embedded in NFW haloes are very resilient to tidal effects until they are nearly destroyed. This is supported by \citet{Navarro10} who found that the NFW shape holds reasonably well even for subhaloes. To roughly gauge the problem, we fitted our (SPH) subhaloes to a NFW profile, and observed that while some of them are well fitted, there are still objects whose density profile cannot be approximated by the simple NFW functional form. Taking all these considerations into account suggests that the NFW profile used to calculate the allowed region is likely not the best choice.

The interpretation of the results presented here clearly demands a closer investigation of the evolutionary tracks of the satellites, the actual influence of the SF and feedback model as well as an improved calculation of the observational confidence level, verifying the applicability of the NFW approach. However, we leave this to a companion paper (Di Cintio et al., in preparation) and only highlight here that simply the inclusion of baryonic physics does not solve the problem; it rather poses new challenges to be explored and studied in greater detail.

\section*{Acknowledgements}
The simulations were performed and  analyzed at  the Leibniz Rechenzentrum Munich (LRZ) and at the Barcelona Supercomputing Center (BSC). We thank DEISA for giving us access to computing resources in these centers through the DECI projects SIMU-LU and SIMUGAL-LU. We also acknowledge the MultiDark  Consolider project and the ASTROSIM network of the ESF for the financial support of the workshop "CLUES workshop" held in Brighton in June 2011, where this paper has been finished. YH has been partially supported by the Israel Science Foundation (13/08). NIL is supported by a grant by the Deutsche Forschungs Gemeinschaft. AK is supported by the {\it Spanish Ministerio de Ciencia e Innovaci\'on} (MICINN) in Spain through the Ramon y Cajal programme and the grants AYA 2009-13875-C03-02, AYA2009-12792-C03-03, CSD2009-00064, CAM S2009/ESP-1496. We like to thank the referee Erik Tollerud for constructive comments.

\bibliographystyle{mn2e}
\bibliography{archive}

\bsp

\label{lastpage}

\end{document}